\documentclass[fleqn,usenatbib]{mnras}

\usepackage{newtxtext,newtxmath}

\usepackage[T1]{fontenc}

\DeclareRobustCommand{\VAN}[3]{#2}
\let\VANthebibliography\thebibliography
\def\thebibliography{\DeclareRobustCommand{\VAN}[3]{##3}\VANthebibliography}

\usepackage{graphicx}	
\usepackage{amsmath}	
\usepackage{color}
\usepackage{xcolor}

\title[Cluster finding for lensed transient discovery]{Enabling discovery of gravitationally lensed explosive transients: a new method to build an all-sky watch-list of groups and clusters of galaxies}

\author[D. Ryczanowski et al.]{
Dan Ryczanowski,$\!^1$\thanks{E-mail: danr@star.sr.bham.ac.uk (DR)}
        Graham P. Smith,$\!^1$
        Matteo Bianconi,$\!^1$
        Sean McGee,$\!^1$
        Andrew Robertson,$\!^2$
        \newauthor{ Richard Massey,$\!^3$
        Mathilde Jauzac $\!^{3,4,5,6}$}
\\
$^1$School of Physics and Astronomy, University of Birmingham, Edgbaston, Birmingham, B15 2TT, UK\\
$^2$Jet Propulsion Laboratory, California Institute of Technology, 4800 Oak Grove Dr., Pasadena, CA 91109, USA\\
$^3$Institute for Computational Cosmology, Durham University, South Road, Durham DH1 3LE, UK\\
$^4$Centre for Extragalactic Astronomy, Durham University, South Road, Durham DH1 3LE, UK\\
$^5$Astrophysics Research Centre, University of KwaZulu-Natal, Westville Campus, Durban 4041, South Africa\\
$^6$School of Mathematics, Statistics \& Computer Science, University of KwaZulu-Natal, Westville Campus, Durban 4041, South Africa
}

\date{Accepted XXX. Received YYY; in original form ZZZ}

\pubyear{2022}

\begin{document}
\label{firstpage}
\pagerange{\pageref{firstpage}--\pageref{lastpage}}
\maketitle

\begin{abstract}
Cross-referencing a watchlist of galaxy groups and clusters with transient detections from real-time streams of wide-field survey data is a promising method for discovering gravitationally lensed explosive transients including supernovae, kilonovae, gravitational waves and gamma-ray bursts in the next ten years. However, currently there exists no catalogue of objects with both sufficient angular extent and depth to adequately perform such a search. In this study, we develop a cluster-finding method capable of creating an all-sky list of galaxy group- and cluster-scale objects out to $z\simeq1$ based on their lens-plane properties and using only existing data from wide-field infrared surveys such as VHS and UHS, and all-sky \textit{WISE} data. In testing this method, we recover 91 per cent of a sample containing known and candidate lensing objects with Einstein radii of $\theta_E \geq 5\arcsec$. We also search the surrounding regions of this test sample for other groups and clusters using our method and verify the existence of any significant findings by visual inspection, deriving estimates of the false positive rate that are as low as 6 per cent. The method is also tested on simulated Rubin data from their DP0 programme, which yields complementary results of a good recovery rate of $\gtrsim 80$ per cent for $M_{200}\geq7\times10^{13}$M$_\odot$ clusters and with no false positives produced in our test region. Importantly, our method is positioned to create a watchlist in advance of Rubin's LSST, as it utilises only existing data, therefore enabling the discovery of lensed transients early within the survey's lifetime.
\end{abstract}




\section{Introduction}\label{sec:intro}

The study of strongly-lensed explosive transients is an emerging field with many applications for modern astronomy and cosmology (see \citealt{Oguri2019} for a review). The magnification associated with gravitationally-lensed sources acts as a free telescope upgrade, allowing us to observe objects that are fainter and farther from us than would normally be possible. Strong lensing also produces multiple images of the same source, which is valuable if the multiply-imaged source is an explosive transient (a short-timescale and non-repeating transient -- hereafter just referred to as a ``transient''), because it can be used to further constrain the mass distribution of the lens \citep[e.g.][]{Sharon15} and precisely measure the Hubble constant (or other cosmological parameters) independently of the cosmic distance ladder \citep{Refsdal64,TreuMarshall16}. In addition, there are currently very few confirmed detections of multiply-imaged transients -- of which all are supernovae (SNe). Within the next decade, we anticipate expansion to other flavours of transient, including gravitational waves and their kilonova (KN) counterparts \citep{GPS19,GPS22}, as well as gamma ray bursts \citep{Ahlgren20}. Not only this, but lensed transient detections are expected to increase vastly in number thanks to deep wide-field surveys such as Rubin's LSST \citep{LSST,OguriMarshall2010,GoldsteinLSNeRates,Wojtak19}, which is expected to discover millions of SNe and thousands of KNe over its lifetime \cite[see table 8.2 in ][]{LSST} -- a fraction of which will be lensed.

The first resolved and spectroscopically confirmed strongly lensed transient (dubbed SN Refsdal) was discovered by \citet{Kelly15} during Hubble observations of galaxy cluster MACS\,J1149.6$+$2223, in which one of the cluster members had lensed a Type-II SN (hosted in a background galaxy at $z=1.49$, \citealt{Smith2009}) into a four-image Einstein cross configuration. The modelling of this cluster as a lens provided strong evidence for the existence of two additional images of SN Refsdal -- one that should have appeared about 10 years prior to this discovery, and one predicted to appear up to around two years in the future \citep{Treu16,Jauzac16}. This future image was indeed detected in concordance with these forecasts about a year later \citep{Kelly16}, enabling measurements of the Hubble constant by several groups \citep{Vega-Ferrero18,Grillo20}.

The second resolved multiply-imaged SN, iPTF16geu, was of particular interest firstly because it was discovered within the data stream sourced from a wide-field survey, and secondly for its classification as a Type Ia {\citep{Goobar17}}. SN Ia have standardisable luminosities, meaning it was possible to more precisely estimate the associated lensing magnification. {However, this SN could not be used for estimating cosmological parameters as this requires a measurement of the time delays between the appearance of successive images of the lensed SN (see \citealt{TreuMarshall16} for a review). In general, time delays caused by isolated galaxy lenses (such as iPTF16geu) are shorter than for objects lensed by clusters, and so require a much greater survey cadence in order to precisely capture when new images arrive. Because of its short time delay, all four images of iPTF16geu were discovered simultaneously, no time delay measurement could be made and therefore no cosmological parameter inference could be completed \citep{More17}. Another similarly highly magnified SN Ia has recently been discovered with a very short time delay \citep{Goobar22}.}

An intriguing candidate multiply imaged SN, AT2016jka \citep[or SN Requiem,][]{Rodney21}, has a much longer predicted time delay of $\sim20$ years between its first and final image. This final image of the SN is predicted to appear in an image of the lensed host galaxy close to the core of the cluster lens. Any possible future estimates of $H_0$ from this system would benefit from the smaller fractional uncertainty achievable from such a long arrival time difference. More broadly, lensed transients with more precise and accurate measurements will help to suppress the uncertainties in $H_0$ obtained from time delay cosmography \citep[e.g.][]{BirrerTreu21}.

In addition to these, many singly-imaged -- but gravitationally magnified -- lensed SNe have also been reported in the literature \citep{Goobar09SinglImgSN,Amanullah11SinglImgSN,clashSingleImgSN,Rodney15SingImgSN,Rubin18SinglImgSN}. Detections of these objects still benefit from lens magnification  
and can still be used to help constrain models of their lens, however single images do not enable measurements of cosmological parameters.

One observing strategy for finding lensed transients follows a watchlist-based approach, whereby wide-field survey telescopes routinely scan the sky for transient events and compare their locations with a list of lens coordinates. {A complete lens watchlist would encompass as many potential lenses as possible across the entire mass range from massive galaxies to clusters, and should span the entire sky in order to maximise prospects of finding lensed sources.}
{However, previous contents of potential watchlists are limited by either the footprint of particular surveys or to lenses selected according to source-plane selection methods -- i.e. objects identified from magnitude-limited searches for arc features. This motivates the need for a more general method to identify potential lenses, in order to populate a lensed transient watchlist.}

When transient detections have been made sufficiently nearby a lens within the watchlist, they are flagged for further investigation and follow-up observations. This technique works to filter out candidate lensed transients from the increasingly large number of events detected nightly by wide-field surveys like those conducted currently by the Zwicky Transient Facility \citep[ZTF,][]{ZTF}, and that will soon commence with Rubin's LSST. 
In addition, the watchlist method allows lensed transients to be identified even if they have multiple images arriving on timescales shorter than the discovery survey's cadence, or if multiply-imaged transients have only one detectable image due to variances in magnification. This is advantageous, as lensed transients with these properties would evade detection by a search that instead looks for multiple images by way of detecting spatially coincident transient events within a wide-field survey that are separated by some time delay.

An important requirement of a lens watchlist is that it contains objects across the full sky. This is because many transients are now being discovered by facilities that are able to monitor the entire celestial sphere, including gravitational wave interferometers and gamma ray burst satellites. This is in addition to the optical surveys that have been, and will continue to monitor a significant fraction of the sky like ZTF and Rubin's LSST. Therefore, current catalogues of lenses that only reside in the footprints of particular surveys are insufficient for maximising lensed transient discoveries. In other words, a significant fraction of the mass function covering the range $10^{12} \lesssim M_{200} \lesssim 10^{15}\,\rm M_\odot$ \citep{Robertson2020} is missing from the tools used to find lensed transients due to the lack of an all-sky lens catalogue. This is especially important in the southern hemisphere, which Rubin will begin to survey in the next few years to unprecedented depths. Ensuring such a list of lenses is available as soon as Rubin's operations begin is ideal so that searches for lensed transients can be optimised immediately -- maximising early science prospects and the baseline over which these discoveries can be made.

It is also crucial to emphasise that the dark matter halos that are efficient lenses for the population of transients in question do not have to have been previously identified as lenses (for example, by identification of arcs or multiple images) to be considered a valid entry in a watchlist. 
This is because, in general, the detectability of lensed transients is independent of whether or not the lensed galaxies that host them can be detected in magnitude-limited surveys -- even optical transients such as SNe can easily be brighter than their host galaxy, so host detection does not always come in tandem.
Therefore, the selection of objects for a complete strongly-lensed transient watchlist should not just consist of known lenses, but in fact any object capable of lensing. In other words, lenses should be selected in the lens-plane (i.e. based on their lensing ability), rather than in the source-plane (i.e. based on a chance alignment with source-plane objects detected in magnitude-limited searches for arcs). As shown in \citet{Ryczanowski}, for the case of group and cluster scale objects, $\sim95$ per cent of $10^{14}$M$_\odot$ clusters and $\sim40$ per cent of $10^{15}$M$_\odot$ clusters would not be identified as lenses based on source-plane selection at the sensitivity of Rubin's early data releases, despite being capable lenses of transients. Therefore, even with next-generation surveys, source-plane lens selection methods will miss a significant fraction of the objects capable of lensing transients. Alternative lens-plane selection methods have been explored previously, \citep[e.g.][]{Wong13,Easycritics19Stapelberg}, {however a catalogue of these objects does not yet exist that fulfils both criteria to be all-sky and sufficiently deep to include the majority of the lenses associated with the high-redshift population of lensed transients.}

Before discussing lens-plane selection methods further, it is first important to consider what objects should populate the watchlist.
Ray tracing through hydrodynamical simulations of large scale structure in the Universe indicates that the optical depth to high magnification ($|\mu|\geq10$) spans a broad range of halo masses, with $\sim50$ per cent of it contributed by objects of group and cluster mass scales ($M_{200}\gtrsim10^{13}M_{\sun}$, \citealt{Robertson2020}). This is a key region of parameter space because at fixed lens magnification clusters produce longer and more precisely measured arrival time differences than galaxy-scale lenses due to the former's larger Einstein radii and denser and flatter density profiles \citep{GPS22}. It is also relatively under-explored in relation to very wide-field time domain surveys, with search strategies for lensed supernovae concentrating on watchlists comprising individual galaxy lenses \citep[e.g.][]{GoldsteinNugent17}.

Previous studies aiming to detect galaxy clusters have historically used a variety of methods. Direct detection is possible through detection of X-rays emitted by the hot intra-cluster gas \citep[e.g.][]{XXL16,Adami18,Koulouridis21}, or distortion of the cosmic microwave background by the same medium -- known as the thermal Sunyaev-Zeldovich effect \citep{tSZ72,PlanckClusts16}. Clusters can also be discovered through observations of galaxies by the detection of a cluster red sequence \citep[e.g.][]{redmapper14}, or by utilising galaxy positions and photometric redshifts to test for clustering in 3-dimensional space \citep[e.g.][]{Eisenhardt08}. More recently, cluster detection in future wide-field surveys was thoroughly tested using a variety of methods in preparation for the wide survey component that is to be conducted by \textit{Euclid} \citep{EuclidComp}. These methods each presented their own merits and challenges, and overall sported great success.
Inspired by these methods, we set out to develop our own that satisfies our requirements for an all-sky lens-plane selected cluster/group catalogue. The method of \citet{Gonzalez19} was of particular interest due to its small number of fundamental requirements and capability to be used alongside existing all-sky survey data, aiming to utilise all-sky $J$-band data from the UKIRT Hemisphere Survey \citep[UHS,][]{ukirtDye18} in the north and the VISTA Hemisphere Survey \citep[VHS,][]{VHS} in the south, as well as $W1$-band data from the {\it WISE} (Wide-field Infrared Survey Explorer) mission \citep{wiseMission}.

Therefore, in this paper we describe and test a method with the capability to produce an all-sky watchlist of galaxy group and cluster-scale objects ($M\gtrsim10^{13}M_{\sun}$) out to $z\simeq1$, with the intention of using the list to aid in the discovery of gravitationally lensed transients. Our method is based upon the principles of \citet{Gonzalez19} and can detect clusters to sufficient depth using only existing all-sky near-infrared data. {The main difference between the method presented here and that of Gonzalez et al. is that our method is tuned to locate groups and clusters out to $z\lesssim1$ for the purpose of populating an all-sky lens watchlist. In Gonzalez et al., the cluster catalogue assembled is limited to the footprints of the surveys used, and focuses on the most massive systems at $z\gtrsim1$, which are less efficient lenses for the predicted populations of transients.
In addition, we formulate a different method of extracting cluster detections and estimating the significance of these detections.}
Furthermore, our testing concentrates on the southern sky which is due to be surveyed by Rubin in the coming years. Prioritising this region allows a curated watchlist to be available to find lensed transients as soon as Rubin begins operations in $\sim2024$.

The structure of this paper is as follows: Section~\ref{sec:survey_data} gives an overview of the surveys and data used in this study. Section~\ref{sec:method} describes the cluster-finding method, and Section~\ref{sec:testing} explains how the method was tested, using both real data and state-of-the-art simulated data, before summarising in Section~\ref{sec:summary}. Where relevant, and unless otherwise stated, we have assumed a flat cosmology with $H_0=67.74$ km/s/Mpc, $\Omega_{\Lambda}=0.693$ \citep{PlanckCosmo15} and give magnitudes in the Vega system.

\section{Surveys and Data}
\label{sec:survey_data}
\subsection{Overview of Surveys}
The main data used in this study are from wide-field surveys conducted with the VISTA and {\it WISE} instruments. These surveys are used due to their complete coverage in the southern hemisphere -- the region due to be surveyed by the Rubin Observatory once in operation -- and also for their access to the wavebands sensitive to galaxies in cluster environments out to $z\simeq1$. Specifically, we make use of \textit{J}-band photometry in the southern hermisphere from VISTA and all-sky \textit{W1}-band photometry from {\it WISE} in order to create maps of the number density of galaxies on the sky. It should be noted that suitable \textit{J}-band data are also available for the northern hemisphere from the UKIRT Hemisphere Survey (UHS), but we prioritise testing in the southern hemisphere due to slightly better magnitude depths and to cover the field of Rubin's LSST. Sources are first matched between the {\it WISE} and VISTA catalogues, giving multi-band coverage of each galaxy -- this ensures all detections are robust, and allows estimates of further properties from $J-W1$ colours such as the redshift. The following sections describe the data sets in more detail.

\subsection{VISTA Data}
The Visible and Infrared Telescope for Astronomy (VISTA) is a 4.1m survey telescope located at Paranal Observatory, Chile. It has conducted a variety of wide-field surveys in the sky above the southern hemisphere and equator, and between these the entire southern sky has collectively been surveyed. In this work, we directly use data from the largest of these surveys: the VISTA Hemisphere Survey (VHS, $\sim$20000 deg$^2$, \citealt{VHS}) and the VISTA Kilo-Degree Infrared Galaxy Survey (VIKING, $\sim$1500 deg$^2$, \citealt{viking}). These surveys provide complete coverage in two near-infrared bands, \textit{J} and \textit{Ks}, with additional coverage in other bands in specific regions. In creating our maps, we utilise the \textit{J}-band (1.25$\mu$m) data which has a 5$\sigma$ detection limit of at least $J(5\sigma)=20.1$ across the surveys. All VISTA data used is publicly available through the online VISTA science archive.

\subsection{{\it WISE} Data}
\label{sec:WISE}
The NASA \emph{Wide-Field Survey Telescope Explorer} ({\it WISE}) is a space-based 0.4m infrared telescope that surveys the entire sky in four infrared bands, named $W1$ to $W4$ with wavelengths of 3.4, 4.6, 12 and 22 $\mu$m respectively. We use sources detected specifically in the $W1$-band -- the most sensitive of the {\it WISE} passbands -- from the CatWISE2020 catalogue \citep{CatWISE2020}, a compilation of almost 2 billion sources collected by the instrument from its first operation in January 2010 up to December 2018.

\begin{figure}
    \centerline{
    \includegraphics[width=1.1\columnwidth]{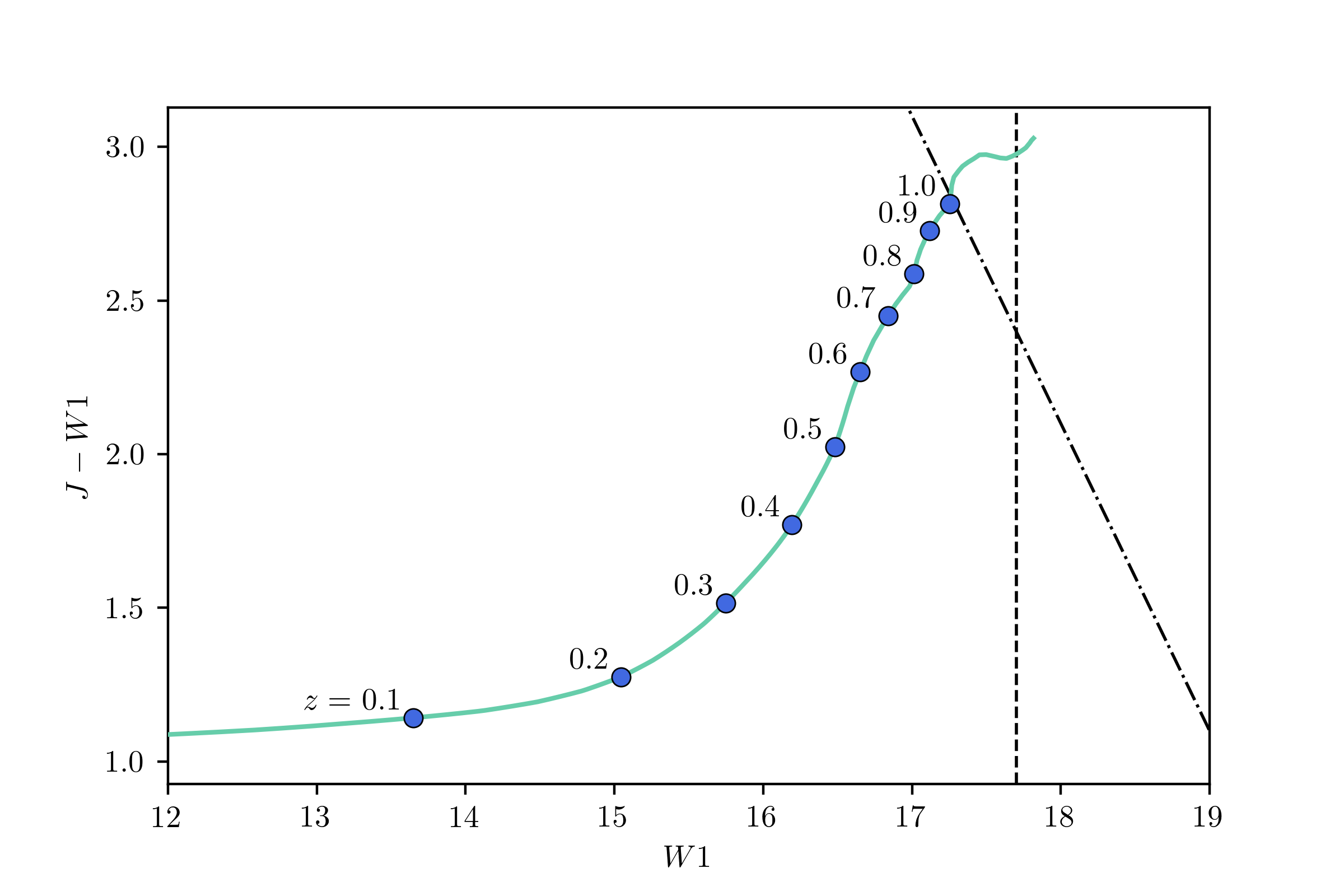}
    } 
    \caption{Predicted $J-W1$ colour evolution of an $L^\star$ cluster galaxy as a function of apparent $W1$ magnitude (blue solid line). orange points highlight the values at specific redshifts during this evolution. The vertical black dashed line represents the 5$\sigma$ $W1$-band magnitude limit of CatWISE ($W1=17.7$), and the dot-dashed diagonal line represents the 5$\sigma$ J-band magnitude limit of VHS ($J=20.1$), the largest of the VISTA surveys. Magnitudes obtained using the EzGal tool assuming a Bruzual \& Charlot evolution, with an SSP star formation history, Chabrier initial mass function and solar metallicity. Given that $z=1$ $L^\star$ cluster galaxies reside within these detection limits, detection of clusters out to this redshift is tractable.}
    \label{fig:depths}
\end{figure}

The addition of the {\it WISE} data is useful for assuring robust galaxy detections by requiring each {\it WISE} source to match to a nearby $J$-band detection from VISTA, and to ensure the galaxies have infrared colours consistent with old stellar populations out to $z\simeq1$.
\autoref{fig:depths} shows the predicted evolution of an $L^\star$ galaxy's\footnote{L* is a characteristic luminosity scale present in the \citet{SCHECHTER} luminosity function, typically representing the value below which the number density grows exponentially.} 
 $J-W1$ colour using the EzGal modelling tool \citep{ezgal} \footnote{\href{http://www.baryons.org/ezgal/}{http://www.baryons.org/ezgal/}}. This shows that with the magnitude limits of each survey, detecting cluster galaxies down to at least $L^\star$ is attainable out to $z\simeq1$. When modelling the galaxy evolution, we assume a Bruzual \& Charlot model \citep{BC03}, with a single delta burst of star formation at $z=3$ that follows a Chabrier initial mass function \citep{chabrierIMF} and contains stars of solar metallicity. We also normalise the luminosity of an $L^{\star}$ galaxy to the \citet{Lin04} cluster sample, which has $K_{\rm S}^\star=15.5$ at $z=0.25$. $W1$ was chosen for this purpose over $K$-band, as $W1$ has full coverage in both hemispheres, and $K$-band in current surveys is not sensitive enough to reach $z=1$.

One important consideration when utilising {\it WISE} data is the large PSF ($W1$-band $\rm FWHM\simeq6\,\rm arcsec$) compared to the VISTA data ($J$-band $\rm FWHM\simeq1\,\rm  arcsec$). This means that in dense cluster cores, there is a possibility that multiple galaxies detected by VISTA are blended into a single {\it WISE} detection. To take account of this, when matching sources between the two catalogues, the number of VISTA sources within one PSF half-width of a {\it WISE} detection is counted and used to weight the contribution of that detection to the galaxy map. The existence of blended sources in dense cluster cores also make it difficult to determine accurate colours for these central galaxies and their neighbours. For example, within our sample of known strong lensing clusters described in \autoref{sec:real_test}, approximately 10 per cent of {\it WISE} galaxies within 1 arcminute of their respective cluster centre have multiple matches in the J-band. However, as we do not make any explicit cuts based on the colours this does not directly affect the density maps, but does have a visible effect on colour-magnitude diagrams we produce during testing in \autoref{sec:false_pos}.

Our selection criteria for galaxies are relatively straightforward, and consist of a flat cut of the 5$\sigma$ detection limit in the relevant bands in both surveys ($J=20.1$ for VHS, the largest of the VISTA surveys and $W1=17.7$ for {\it WISE}) and a cut on the VISTA catalogue's {\it pGalaxy} attribute which estimates the probability that a source is a galaxy. We require {\it pGalaxy} $>0.9$ to eliminate any interloper stars, although the majority of the sample has {\it pGalaxy} $>0.99$, indicating these are secure detections of extended sources.

\section{Method}
\label{sec:method}
\subsection{Background}

Galaxy cluster members are typically early-type galaxies that emit strongly in the near-infrared, due to the spectrum of metal-poor population II stars ($T\simeq3000\rm K$) that dominate early type galaxies' peaks at a rest-frame wavelength of $\lambda\simeq1\mu\rm m$. Utilising the $W1$-band of {\it WISE} and the \textit{J}-band of VISTA, we produce density maps of galaxies detected in both catalogues. This is done following a similar methodology to \citet{Gonzalez19}, whereby a raw map of galaxy positions is convolved with a difference-of-Gaussians smoothing kernel, {but we use slightly different data sets to produce the maps}. This convolved map then provides an estimate of the local density of galaxies within that region of sky, and peaks within the map can hence be used to unveil candidate galaxy groups and clusters within the wide-field data. {We then develop a new method to estimate the significance of peaks referring to cluster detections within the maps.} 

Given that no specific colour or redshift information is taken into account when selecting galaxies, it is entirely possible that multiple chance alignments of smaller groups can produce signals in the maps comparable to those from richer clusters. Whilst it might initially appear that {such detections} are problematic, such cases are still considered to be high-density lines of sight and are therefore still valuable from a strong lensing perspective, and hence belong in a lensed transient watchlist. Therefore, we can motivate using this method to assemble a list of the densest regions of the sky as traced by collections of galaxies detected in infrared data. {Furthermore, we are more concerned with the fast creation of a watchlist in advance of Rubin, which can then be refined based on specific halo properties once higher-quality data are obtained following commencement of the LSST. These lines of sight may not contribute the same lens potential as a singular massive halo, but filtering them would risk reducing the number of interesting lines of sight too significantly -- of the 270 serendipitous detections used for estimating false positive rates in \autoref{sec:false_pos}, 40 ($\sim15$ per cent) show two or more distinct photometric redshift peaks, indicative of multiple halos aligned along the line of sight. Identification of such cases also requires additional photometric redshift data that are not available all-sky, so removing these detections large-scale is not feasible with current data.}

\subsection{Creating density maps with kernel convolution}
\label{sec:convolution}
To create a raw density map, galaxies matched between the two catalogues that pass the selection criteria are placed into a $0.5\times0.5$ degree pixel grid with 7.5 arcsec/pix scale. {This is almost identical to the approach used by \citet{Gonzalez19}, except using a slightly different pixel scale and we assign our maps to cover smaller regions, allowing for multiple disconnected regions to be mapped. {This is done specifically so we can test the method on smaller regions containing known group and cluster-scale objects.}} Each pixel contains a value representing how many galaxies exist within that region of space, and we take into account the large PSF of {\it WISE} by weighting the contribution of a single {\it WISE} detection by the number of \textit{J}-band detections from VISTA within the {\it WISE} PSF radius. This ensures we are not underestimating the number density due to blended sources in dense cluster cores where there can be up to of order 10 blended galaxies.

Once the raw map has been constructed, it is convolved with a kernel designed to smooth out contributions from components that do not match those of typical cluster core scales. The kernel is described by a difference of two normalised Gaussian functions with different widths given by:
\begin{figure}
    \centerline{
    \includegraphics[width=1.0\columnwidth]{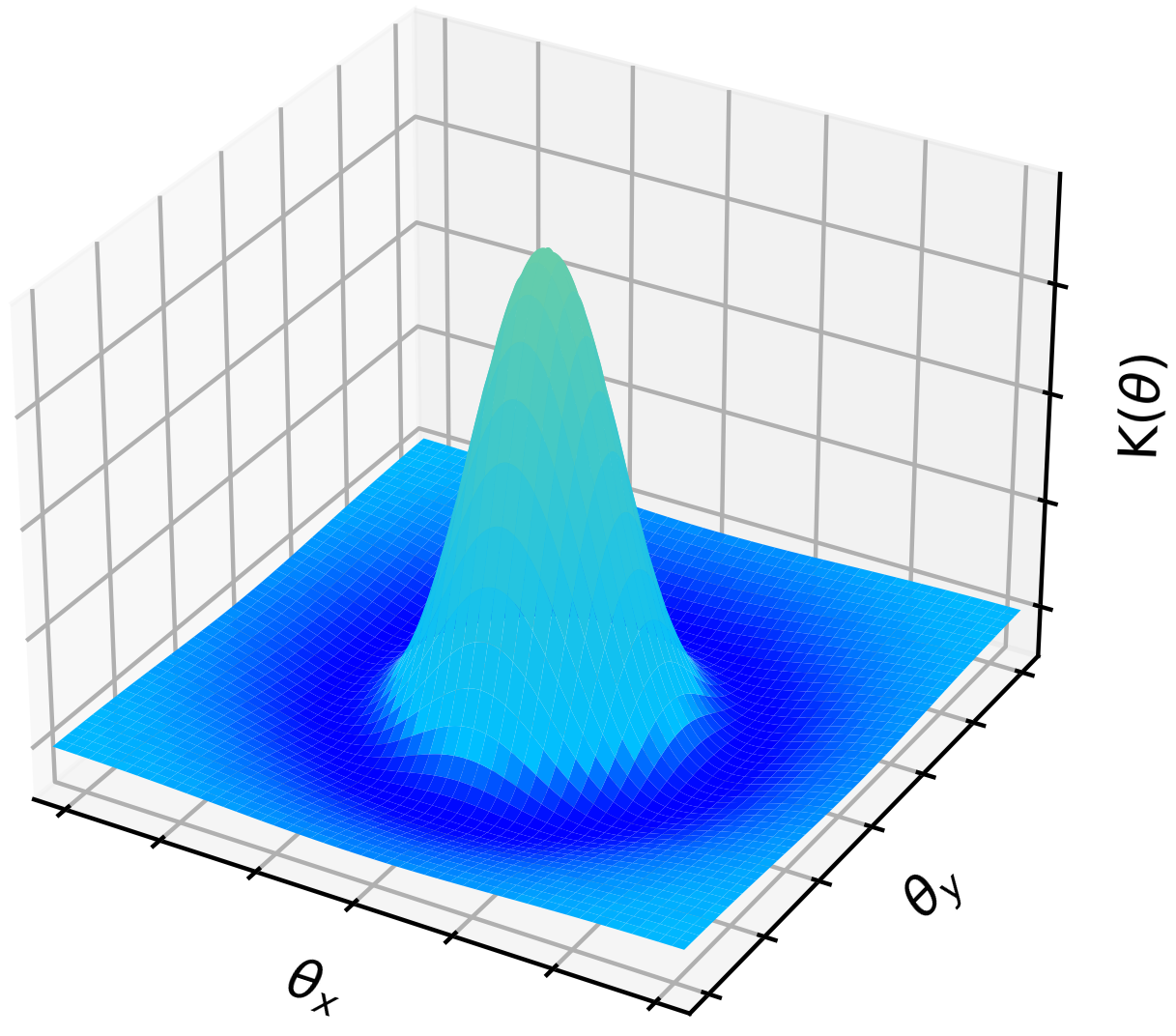}
    }
    \caption{Example surface plot of a normalised difference-of-Gaussians kernel, showing the positive central peak and the negative region surrounding it. This is used as a smoothing function to identify cluster-like structures within pixellated galaxy maps. The angular extent of the interior width is tuned to that of cluster cores, and acts to smooth out contributions from scales larger and smaller than this.}
    \label{fig:diff_gauss}
\end{figure}
\begin{equation}
    K(\theta) = \frac{1}{2\pi\sigma_{\textrm{in}}^{2}\sigma_{\textrm{out}}^2}  \left[\sigma_{\textrm{out}}^2\exp\left(\frac{-\theta^2}{2\sigma_{{\textrm{in}}}^2}\right)-\sigma_{\textrm{in}}^2\exp\left(\frac{-\theta^2}{2\sigma_{\textrm{out}}^2}\right)\right],
    \label{eqn:diffGauss}
\end{equation}
where $\theta$ is the angular separation and $\sigma_{\textrm{in}}$, $\sigma_{\textrm{out}}$ are the widths of the inner and outer Gaussians, respectively. By definition, $\sigma_{\textrm{in}}$ < $\sigma_{\textrm{out}}$. Each Gaussian is normalised to have unit volume, such that the kernel integrates to zero. Similarly to \citet{Gonzalez19}, we select $\sigma_{\textrm{in}} = 45$ arcsec, corresponding to the scale of the dense cores of galaxy clusters, and choose the ratio of $\sigma_{\textrm{in}}$ to $\sigma_{\textrm{out}}$ to be 1:6. Multiple values were experimented with for the filter widths, anticipating an effect based on the changing scales of clusters at different redshifts -- however the final results were insensitive to any change of these parameters. \autoref{fig:diff_gauss} shows an example of a difference-of-Gaussians function as given by \autoref{eqn:diffGauss}.

\begin{figure}
    \hspace{-5mm}
    \includegraphics[width=1.1\columnwidth]{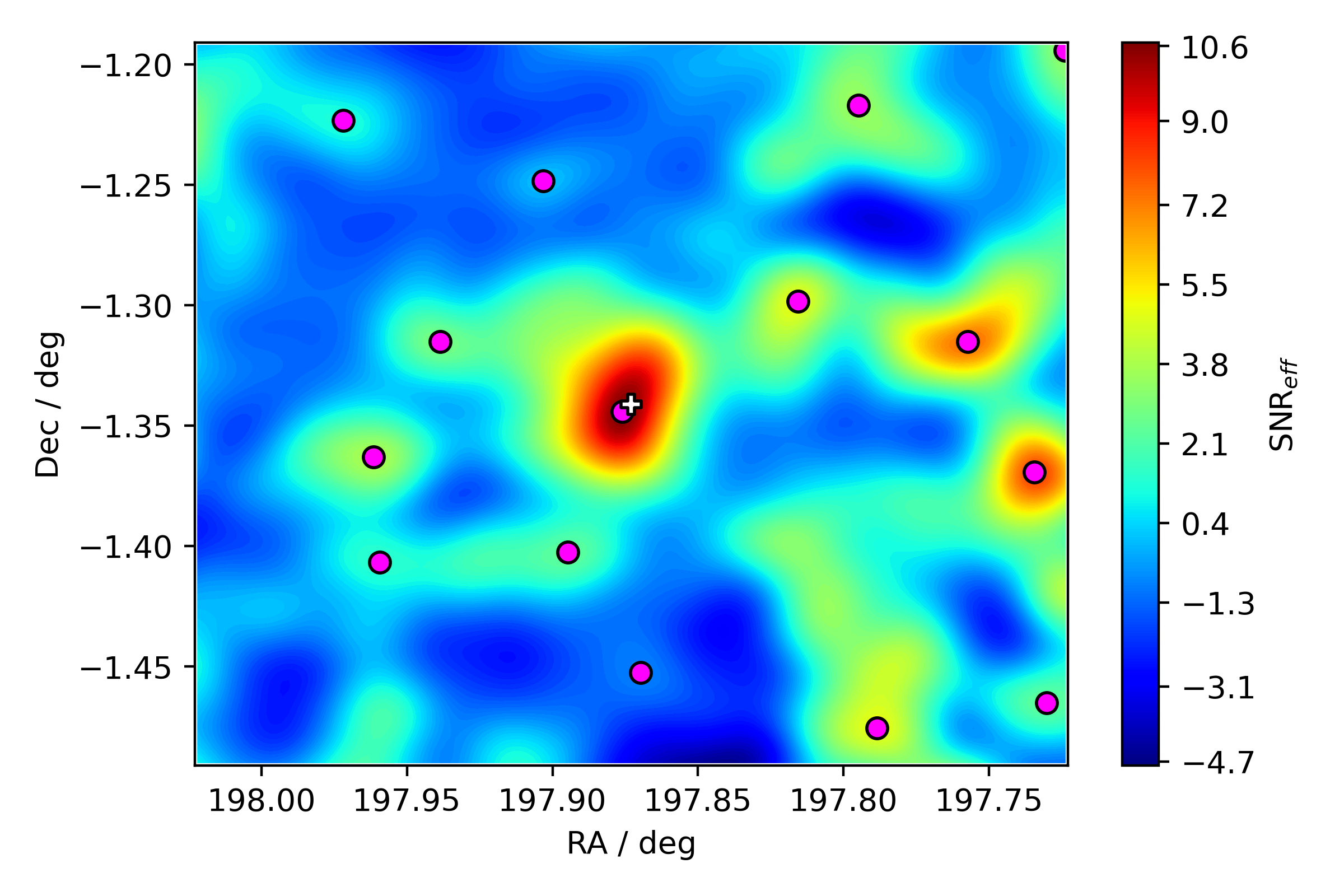}
    \caption{A $0.3\times0.3$ degree map produced by convolving a difference-of-Gaussians kernel with a pixel grid of galaxy positions surrounding the coordinates of known cluster and strong lens Abell 1689. Purple dots highlight peaks in the galaxy number density distribution, and the white plus marks the coordinates of the cluster centre, which in this case is located very close to the largest peak. The colour bar scale shows the SNR$_{\rm eff}$ of each pixel, a quantity introduced in \autoref{sec:SNR} which indicates the significance of a detection, and is normalised between the maximum and minimum values within the map. Data to produce this map are taken from an overlapping $0.5\times0.5$ degree region, extending an additional 0.1 degrees from each edge of the map. This is to ensure there is sufficient data beyond the map region to prevent any edge effects caused by the convolution. The cluster in the centre is recovered with high significance, and a few other regions are highlighted as having a high density along the line of sight and hence represent candidate groups/clusters.}
    \label{fig:density_map_example}
\end{figure}

Applying the convolution produces a map whose pixels correspond to the local galaxy number density; \autoref{fig:density_map_example} shows an example map centred on known galaxy cluster and strong lens Abell 1689.

\subsection{Map size and edge effects}
\label{sec:mapsize}
Within each density map, a peak-finding algorithm is utilised to identify positions of the densest regions of the map. To reduce the impact of edge effects caused by the convolution on outer regions of a map, we ignore any peaks that are within 24 pixels of the map border. This effectively reduces the size of each map to $0.3\times0.3$ deg, but ensures that no bias is introduced to outer regions. {The reduced size of map is a fairly arbitrary choice that has been seen to work well on test regions containing known objects -- the average pixel values around the edge of the maps are not significantly different to typical values closer to the centre, signifying that the edge effects are not prominent. The (reduced) map size can be optimised for efficiency in a blind search by determining the maximal usable area for a given map that is not noticeably affected by any edge effects. This can be done by looking at how the average pixel value changes in successively larger annuli further from the centre of the post-convolution map, and choosing a suitable tolerance for how much edge effects can vary pixel values}, but we leave the investigation of this for future work.

The limits on the size of a map are based only on the memory limits of the machine processing the data. We chose to use the size we did to efficiently create maps for many disconnected regions {containing known group and cluster-scale objects, which is efficient for testing.} However, a blind search over a larger continuous footprint could in theory be done as a smaller number of much larger maps on a more powerful machine, which would significantly reduce the fraction of regions affected by edge effects.

\subsection{Estimating overdensity significance}
\label{sec:SNR}

{Once the peak-finding algorithm has been run on the reduced map,} we quantify the significance of each overdensity in a map. This is done by adopting a quantity similar in form to a signal-to-noise ratio (SNR), which we call the ``effective SNR'', SNR$_{\textrm{eff}}$, and calculating this for each pixel in the convolved map. We then take the largest SNR$_{\textrm{eff}}$ pixel value of each local maximum to represent the significance of a detection at that position. {This is a new way to formulate the significance of each candidate cluster.} To calculate this quantity, we first assume that the background of non-clustered galaxies is randomly and uniformly distributed. Under this assumption, we can produce a large set of maps of randomly-distributed galaxies, where each map contains the same total number of galaxies as the real map now randomly positioned across the same size patch of sky. These random maps are then convolved by the same difference-of-Gaussians kernel, and can be used as a metric to estimate the background noise. By calculating the mean and standard deviation of pixels within these random maps, we define SNR$_{\textrm{eff}}$ for each pixel within real maps as:
\begin{equation}
    \textrm{SNR$_{\textrm{eff}}$} = \frac{s-\mu}{\sigma} = \frac{s}{\sigma},
    \label{eqn:SNR}
\end{equation}
where \textit{s} is a pixel value, and $\mu$ and $\sigma$ are the pixel mean and standard deviations respectively, determined from the sample of random maps. By construction, any pixel map convolved with a kernel of the form given by \autoref{eqn:diffGauss} will have $\mu=0$, due to the kernel's property of integrating to zero, hence the additional simplification in the equation and thus leaving only $\sigma$ to be determined. In a random map containing $N_{\textrm{gal}}$ galaxies, the pixel values follow a Poisson distribution and so $\sigma\propto\sqrt{N_{\textrm{gal}}}$. Given that the smoothing kernel acts only to spread the pixel values from a raw map over a larger region, one can expect the standard deviation of pixel values within the entire convolved map to follow the same distribution. Therefore, we can precisely measure the standard deviation $\sigma'$ for a single arbitrary number of galaxies $N_{\textrm{gal}}'$ and calculate SNR$_{\textrm{eff}}$ in proportion for the relevant $N_{\textrm{gal}}$:

\begin{equation}
    \textrm{SNR$_{\textrm{eff}}$} = \frac{s}{\sigma'}\sqrt{\frac{N_{\textrm{gal}}'}{N_{\textrm{gal}}}}.
\end{equation}
Consequently, random maps only need to be created once, with new random maps being required only if a fundamental property of the method, such as the kernel size, is changed. By creating $10^4$ random maps each containing $N_{\textrm{gal}}' = 2000$ galaxies, we determine $\sigma' = (8.7\pm0.4)\times10^{-3}$ galaxies/pixel. These values are then used to determine SNR$_{\textrm{eff}}$ within the pixels of the real maps, which is in turn used to assess the confidence that a given local maximum within a map corresponds to a real group or cluster-scale object, and comparatively rank any peak determined by this method by way of proxy for the mass or richness of the cluster core. We discuss in \autoref{sec:testing} the values of SNR$_{\textrm{eff}}$ that categorise the robustness of detections. It should be stressed that SNR$_{\textrm{eff}}$ is not a true signal-to-noise ratio, but rather an estimator based on correlated nearby pixels that aims to evaluate the overdensity of a region of galaxies under simple assumptions.

Given that the non-clustered background galaxy distribution will not be truly random but rather correlated on small scales to, on average, be denser than random, SNR$_{\textrm{eff}}$ will be a slight underestimate due to underestimating the magnitude of the noise. However, we believe this effect to be small due to the amplitude of the galaxy two-point correlation function at scales similar to those of our maps. Figure 15 in \citet{Wang13_2p_corr_fn} shows a determination of this quantity -- their faint population of galaxies is approximately representative of the typical L* cluster galaxies we expect to be detectable with {\it WISE}, based on colours inferred from the same EzGal models introduced in \autoref{sec:WISE}. Based on their results, galaxies at scales similar to our maps ($\sim$0.1 deg) are $\sim$5 per cent denser than a purely random distribution. Therefore, we expect the difference made due to our simplifying assumptions to be small. In addition, the rank ordering of objects based on SNR$_{\textrm{eff}}$ is arguably more useful than the specific values of SNR$_{\textrm{eff}}$ alone, and this would largely remain unchanged as a result of this assumption.

\section{Testing the method}
\label{sec:testing}
\subsection{Known cluster/group-scale lens test sample}
\label{sec:real_test}
We use two samples of lenses to test our method's ability to recover known groups and clusters -- 130 spectroscopically confirmed cluster-scale lenses assembled by \citet{Smith18}, and 98 galaxy and group-scale lenses assembled by \citet{CarrascoClusterSample}. The cluster-scale lenses have been utilised in previous searches for lensed gravitational waves and supernovae with a prototype watchlist \citep{Smith18lensGWsearch,gps19obs,ryczanowski21_SN_obs,Bianconi22}. These 130 objects are some of the best studied strong lenses in the literature, many of which have been observed by the \textit{Hubble Space Telescope}. This sample ensures our test sample extends to the most extreme clusters in terms of mass, {as the Einstein radii range from $3''\lesssim\theta_E\lesssim60''$, and hence includes individual objects with some of the largest strong lensing cross-sections known.} The distribution of these clusters peaks at $z\sim0.35$, but the upper tail stretches to $z\sim1$, which suitably matches the distributions of objects we aim to detect, as outlined in \autoref{sec:intro}.

Objects from the \citet{CarrascoClusterSample} sample are lenses and lens candidates that were found within the Canada–France–Hawaii Telescope Lensing Survey (CFHTLenS) data. CFHTLenS is a deep optical ($u^*/g'/r'/i'/z'$-band) survey spanning $154\,\rm deg^2$ observed by the Canada-France-Hawaii Telescope (CFHT) as part of the Canada–France–Hawaii Telescope Legacy Survey (CFHTLS). The survey's primary objectives involve studies of weak lensing effects, however the high-quality data along with accurate photometric redshifts makes it useful for some strong-lensing science as well. These group and galaxy scale objects typically have smaller Einstein radii than those in the cluster sample, ranging from $3''\leq \theta_E \leq 18''$, which is valuable for testing the limits of what the method can reliably detect. The redshift range of $0.2<z<0.9$ is also well-matched to our aims and the cluster sample. 

We were able to retrieve VISTA data for 63 out of the 98 objects within the \citeauthor{CarrascoClusterSample} sample. The majority of the objects that do not have VISTA data reside in the third CFHTLenS region around $\delta=55\deg$ -- far outside of the coverage of the surveys. We were also only able to obtain data for 62 out of the 130 sample of known lenses, giving a total of 125 objects in our final test sample. Ultimately, the only information we utilise in our method beyond selection is the coordinates of objects from the samples and use them to obtain VISTA and {\it WISE} data from the surrounding regions. This allows us to test our method fully, extract the SNR$_{\textrm{eff}}$ distribution for regions containing known objects and hence estimate the recovery rate of the method.

\subsection{Results of the known lens test}
\label{sec:results_real}

For each of our sample objects, we produce a convolved density map using the difference-of-Gaussians kernel and the procedure as described in \autoref{sec:convolution}. Within each map, we locate peaks within $5\sigma_{\textrm{in}}$ (225 arcsec) of the object's coordinates given in the sample and calculate SNR$_{\textrm{eff}}$ for each of these peaks using the same method discussed in \autoref{sec:SNR}. If the SNR$_{\textrm{eff}}$ around the peak closest to the object is above some designated threshold (which we initially define to be 5), then the object is considered to be recovered with high significance.

\begin{figure}
\hspace{-5mm}
    \includegraphics[width=1.15\columnwidth]{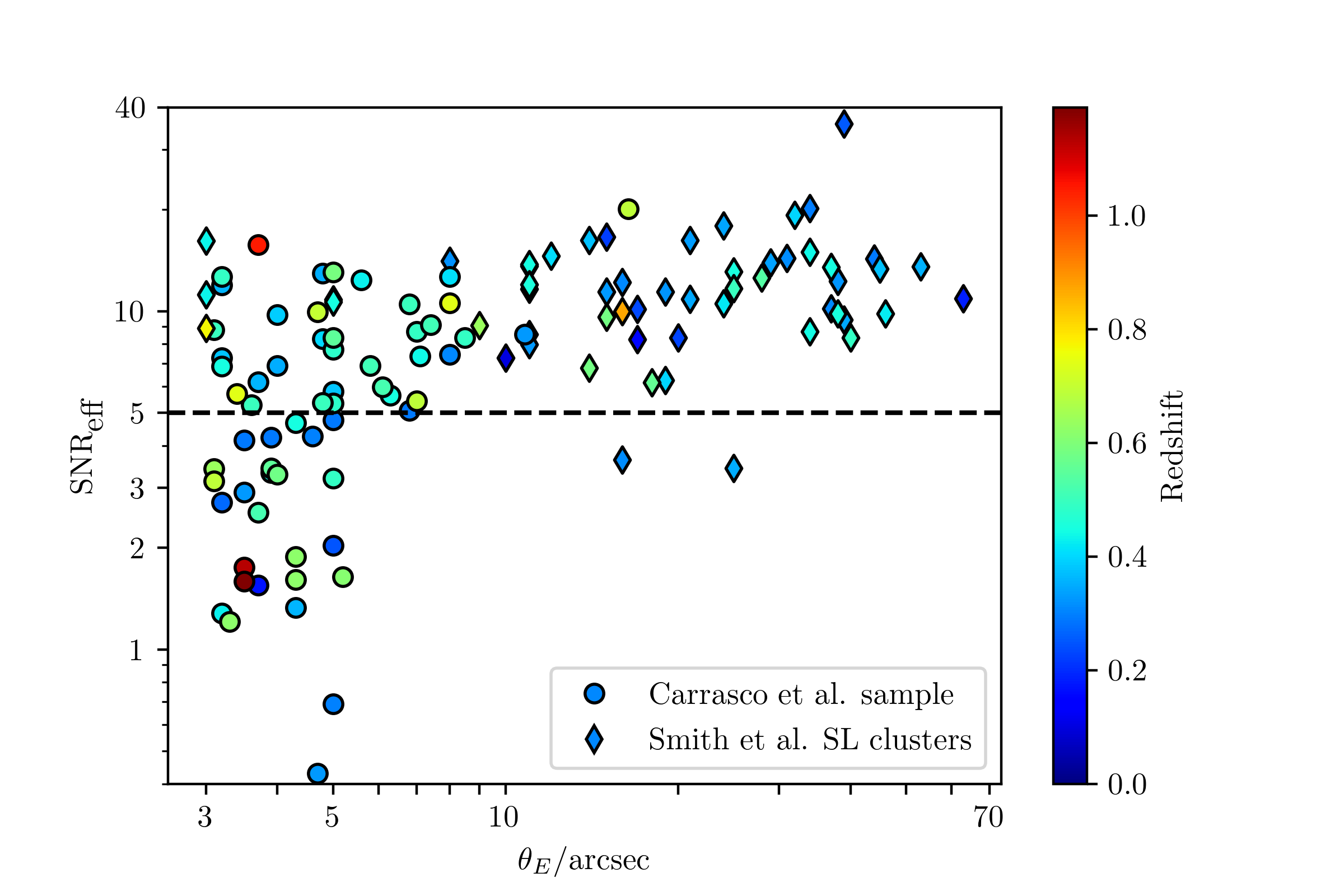}
    \caption{Distribution of SNR$_{\textrm{eff}}$ against Einstein radius for the 125 test objects, made up of those from the \citeauthor{CarrascoClusterSample} sample {(circles) and the \citeauthor{Smith18lensGWsearch} SL sample (diamonds)}. Horizontal dashed line at SNR$_{\textrm{eff}}$=5 marks the threshold above which an object is classed as recovered with high significance. The colour of the points represents the redshift of the object, signified by the colour bar. Recovery appears to be related to Einstein radius, as the majority of non-recovered objects have small $\theta_E$ ($\theta_E\lesssim5\arcsec$). Recovery also does not appear to depend strongly on redshift, as shown by the absence of any obvious trends with redshift.}   \label{fig:snr_erad}
\end{figure}

\begin{figure}
    \hspace{-5mm}
    \includegraphics[width=1.0\columnwidth]{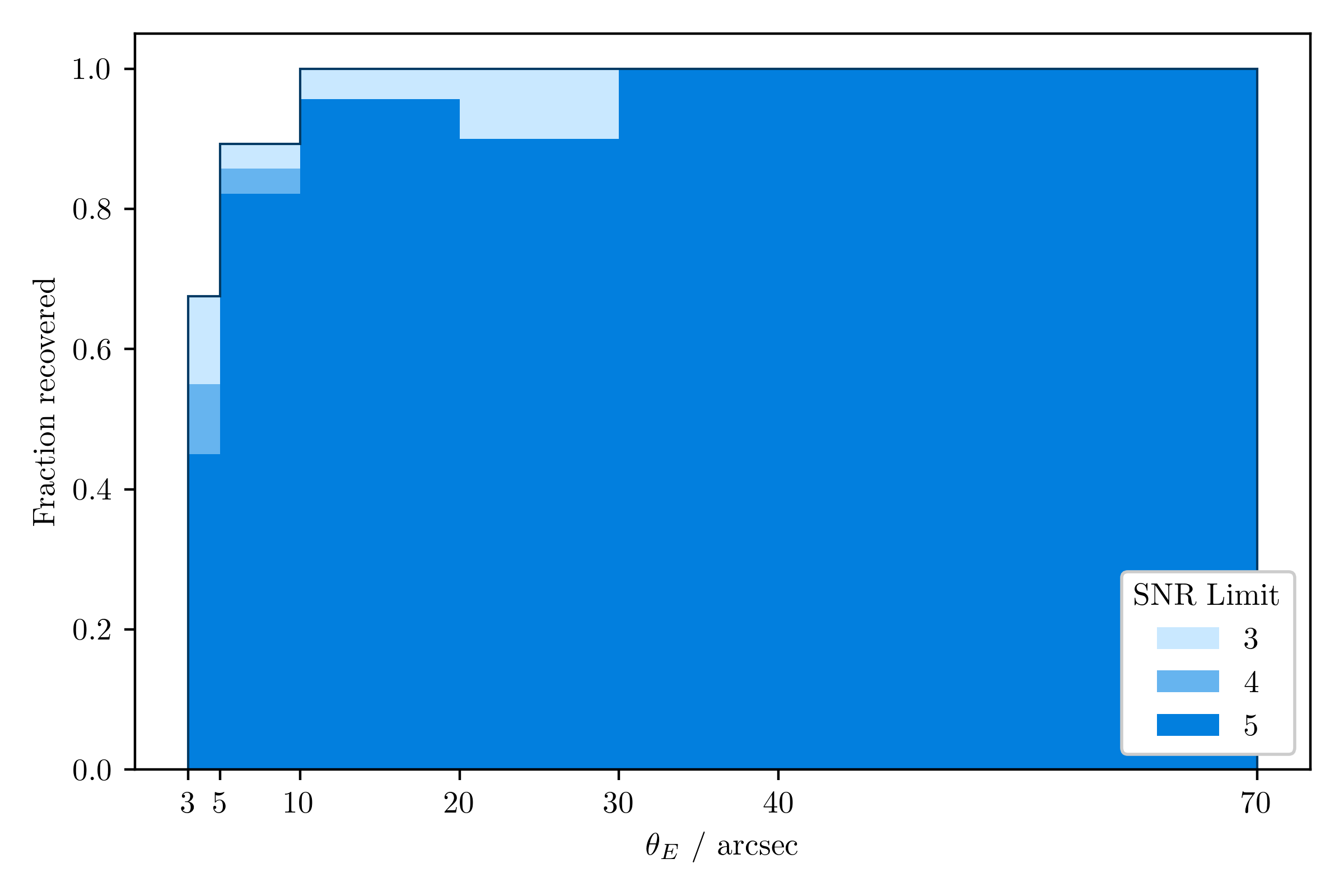}
    \caption{Histogram showing the recovery fraction of objects in bins of Einstein radius. Different colour bars represent different SNR$_{\textrm{eff}}$ thresholds to define recovery.
    Even with the stringent requirement of SNR$_{\textrm{eff}} > 5$, the recovery fraction of our sample is >80\% for $\theta_E>5''$, and 100\% of $\theta_E>30''$ objects are recovered, with an average of 91\% recovered for SNR$_{\textrm{eff}}=5$. The reason for the drop off below 5$\arcsec$ is partially due to the sample containing isolated galaxy lenses as well as smaller galaxy groups, and this method is naturally less sensitive to objects with fewer members. Using a lower SNR$_{\textrm{eff}}$ threshold of 3 allows for complete recovery of all $\theta_E>10\arcsec$ objects, but this will come at the expense of a greater number of false positive detections.}
    \label{fig:recov_frac_histogram}
\end{figure}

\autoref{fig:snr_erad} shows a plot of the calculated SNR$_{\textrm{eff}}$ against estimated Einstein radius, $\theta_{E}$, for our 125 sample objects. It should be noted that the $\theta_{E}$ values for objects in the \citet{CarrascoClusterSample} sample are taken directly from their catalogue and are only estimates found, as they describe in their paper, by finding the average distance between the bright arc (or candidate arc in the unconfirmed cases) and the object's centre. We use $\theta_{E}$ as a proxy for the lensing cross-section (and equivalently, mass) of each object as is commonplace in many lens models. Objects with small $\theta_{E}$ tend to produce smaller SNR$_{\textrm{eff}}$, making up the majority of non-recovered objects with SNR$_{\textrm{eff}}$ $<5$.
{ In addition, above $\theta_E\sim5\arcsec$, there is a slight correlation between Einstein radius and SNR$_{\textrm{eff}}$, with the data in \autoref{fig:snr_erad} showing a Spearman correlation coefficient of $r_S = 0.504$, indicating clusters with larger Einstein radii generally produce slightly higher SNR$_{\textrm{eff}}$. As there is inherent scatter in the Einstein radii for clusters of a given mass or richness in both the \citeauthor{Smith18lensGWsearch} and \citeauthor{CarrascoClusterSample} samples, we would not expect a perfect correlation between the two variables.}
These results are further highlighted by \autoref{fig:recov_frac_histogram}, which shows the fraction of objects recovered in bins of $\theta_E$. The recovery fraction drops significantly for $\theta_E<5\arcsec$, due to the inclusion of single galaxy and small group lenses within the \citeauthor{CarrascoClusterSample} sample, but is much higher (>80 per cent recovery) above this. Given that SNR$_{\textrm{eff}}$=5 is fairly arbitrary, since SNR$_{\textrm{eff}}$ is only an estimator for an object's overdensity, we can vary the cut to see the effect on recovery. A cut at SNR$_{\textrm{eff}}$=3 increases the recovery rate at low Einstein radii fairly significantly, and allows the remainder of the larger radius objects to be recovered, giving 100 per cent recovery for $\theta_E>10\arcsec$. However decreasing the SNR$_{\textrm{eff}}$ threshold will increase the proportion of false positives in a blind search, as we explore in the next section, so this must be done with caution.

The drop in recovery fraction for the $\theta_E<5\arcsec$ objects is unsurprising, as some of these objects are single galaxy lenses or small galaxy groups, which will naturally be either impossible or very difficult to detect by any method sensitive to number density.
It is also interesting to note that \autoref{fig:snr_erad} suggests the redshift of the object does not appear to have a major impact on whether the object is recovered or not. There are objects with a wide range of redshifts both above and below SNR$_{\textrm{eff}}=5$ -- although the impact on SNR$_{\textrm{eff}}$ due to redshift at the upper end of the range ($z\gtrsim0.8$) is not well tested due to the scarcity of these objects. 

\begin{figure}
    \hspace{-5mm}
    \includegraphics[width=1.15\columnwidth]{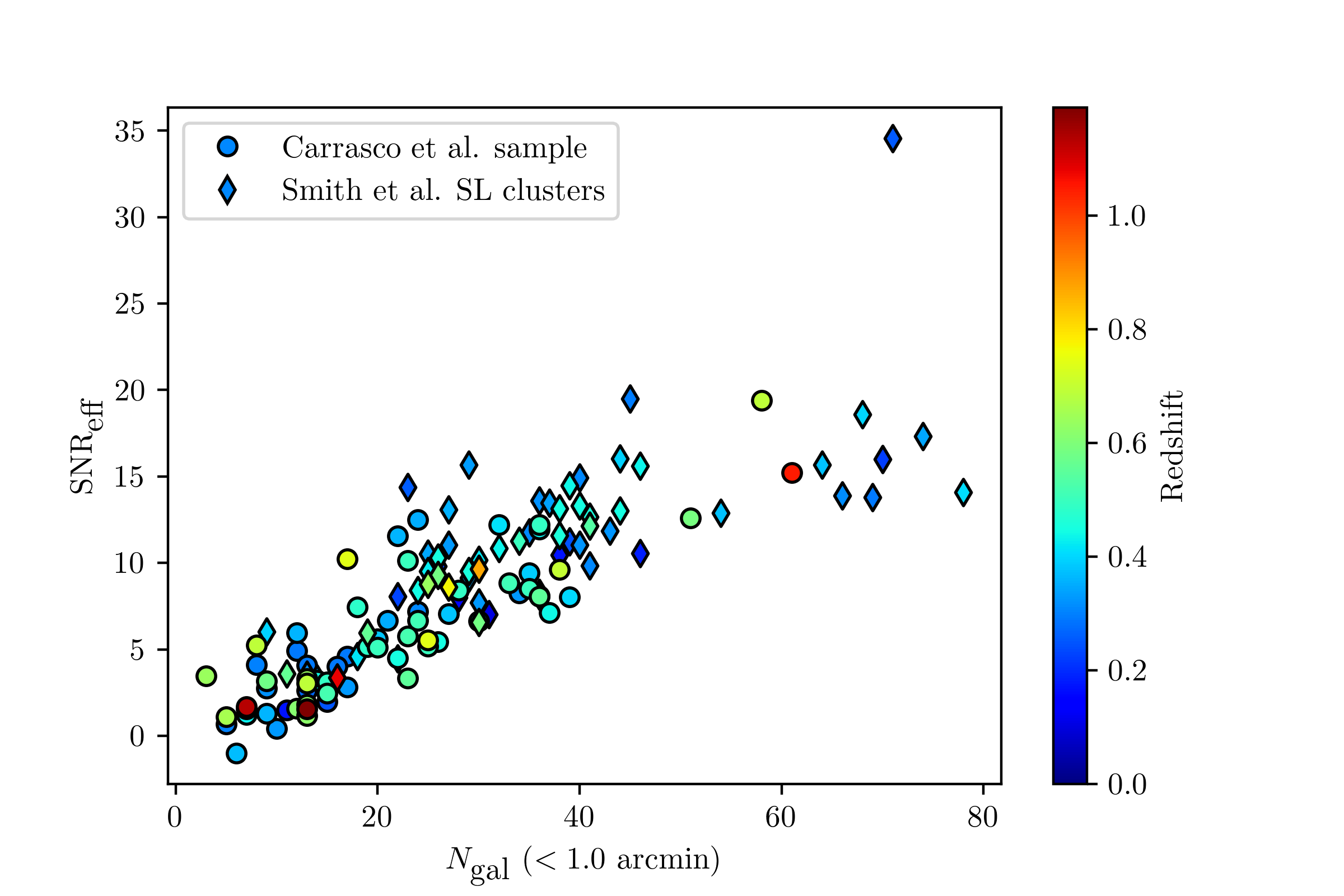}
    \caption{SNR$_{\textrm{eff}}$ of objects in the two test samples against the weighted number of galaxies detected within 1 arcmin of the corresponding map peak. The colour of each point indicates the redshift of the object. The SNR$_{\textrm{eff}}$ appears to correlate with the number of galaxies, although there is some clear scatter present. Therefore, SNR$_{\textrm{eff}}$ is a reasonable estimator for the richness of an object.}
    \label{fig:SNR_ngal}
\end{figure}

\autoref{fig:SNR_ngal} shows the SNR$_{\textrm{eff}}$ against the {weighted} number of galaxies detected in the matched VISTA-{\it WISE} catalogues within 1 arcminute of the map peak closest to the coordinates of each lens in our sample. {As explained in \autoref{sec:convolution}, the weighting is done based on the number of blended $J$-band detected galaxies within the PSF of each \textit{WISE} detection.} This affirms that the method is more sensitive to objects with a larger member number density, given that there appears to be a scattered but approximately linear relationship between the number of selected galaxies within 1 arcmin of the peak and the corresponding SNR$_{\textrm{eff}}$ {(with Spearman correlation coefficient $r_S = 0.870$). We suspect the larger $r_S$ for this relation compared to SNR$_{\textrm{eff}}$ against Einstein radius in \autoref{fig:snr_erad} is because the cluster richness will directly influence the pixel values within the map that drive the SNR$_{\textrm{eff}}$ calculation, and additionally because there is inherent scatter between Einstein radius values for a given cluster mass or richness. The calculations of $N_\textrm{gal}$ here ignore} the fact that each 1 arcminute region will be contaminated by galaxies that are not part of the group or cluster, however by assuming contaminant galaxies are randomly distributed (or close to it), this should only contribute to the scatter which would be overshadowed by the cluster overdensity.

{Within \autoref{fig:snr_erad} and \autoref{fig:SNR_ngal}, there are two clear anomalous data points that stand out from the rest. Firstly is the high redshift $(z\sim1)$ and high SNR$_{\textrm{eff}}$ ($\simeq15$) point. This particular object is a high-redshift group that is closely aligned with a larger and closer cluster ($z\sim0.3$) along the line of sight. Therefore, the detection appears with an inflated $N_{\textrm{gal}}$ and hence SNR$_{\textrm{eff}}$ compared to objects at a similar redshift.
The other is Abell 1835 -- the object at SNR$_{\textrm{eff}} \simeq 35$, which is significantly higher than any other object in our sample, especially considering there are other clusters with similar $N_{\textrm{gal}}$. This object in particular has a large number of highly-weighted galaxies near to the cluster core, hence our method assigns this cluster a high SNR$_{\textrm{eff}}$ even among the large $N_{\textrm{gal}}$ objects.}

\subsection{False positive rate estimates}
\label{sec:false_pos}

Our method is based on counting galaxies without considering their colours. This is well matched to the available all-sky data given that the colour of galaxies in crowded cluster and group cores is challenging to measure accurately given the mis-match in angular resolution between ground-based $J$-band ($\lambda=1.25\mu\rm m$) and space-based $W1$ ($\lambda=3.4\mu\rm m$) photometry. In contrast, modern galaxy-based cluster- and group-finding algorithms typically incorporate colour information \citep[e.g.][]{redmapper14,Gonzalez19}. Objects detected by our method may therefore include projections of more than one group or cluster along the line of sight, and chance alignments of galaxies at very different redshifts that are indistinguishable from real groups or clusters when colour information is not taken into account. Given that our focus is on identifying group and cluster-scale dark matter halos capable of strong lensing, we regard the former (multiple groups and clusters) as a secure detection, and the latter (chance projection of field galaxies) as false positives. 

\begin{figure*}
    \centering
    \includegraphics[width=2.0\columnwidth]{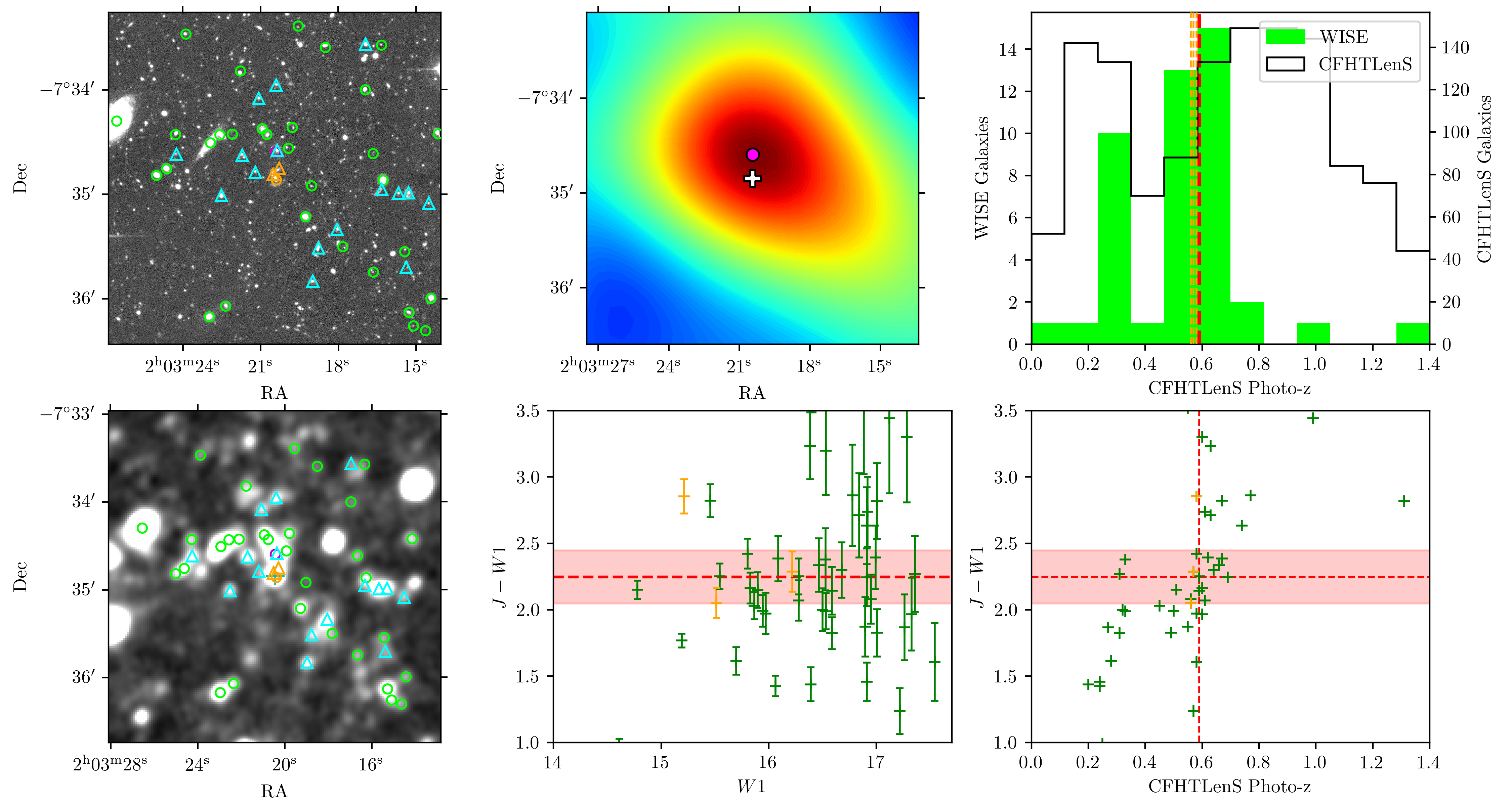}
    \caption{Example of a series of plots used to confirm the existence of objects within our test sample, at the coordinates provided by the original source. This series represents object SA6 from the \citet{CarrascoClusterSample} sample. {Top-left: an image of the region from optical CFHTLenS data showing the galaxies detected within the $J$ and $W1$ data. Top-centre: A zoomed in version of the density map, centred on the largest peak, highlighting the closeness of the peak to the recorded coordinates of the object. Top-right: histograms of the redshifts of galaxies within the CFHTLenS catalogue (black), and a subset of those that also appear in the {\it WISE} data (green). This shows two distinct redshift peaks at $z\sim0.3$ and $0.6$ indicating the presence of real collapsed objects at these redshifts. Bottom-left: same as top-right, except using {\it WISE} imagery. Bottom-centre: a $J$-$W1$ colour-magnitude diagram for the galaxies within this region, showing a cluster red sequence in accordance with the predicted colour for the object. Bottom-right: colour against photometric redshift for galaxies in the colour-magnitude diagram. This shows a cluster of points around the predicted colour and recorded photometric redshift of this object, furthering the evidence for a real object at this location.}}
    \label{fig:6panel}
\end{figure*}

We estimate the false positive rate from the population of detections in the sky regions surrounding the known lenses from the \citeauthor{CarrascoClusterSample} sample. Specifically, we check whether there are obvious peaks in the CFHTLenS photometric redshift catalogue associated with detections other than the known lens in our $0.3\times0.3$ degree $J/W1$-based maps centred on these lenses. In practice, this involved inspecting figures following the same format as \autoref{fig:6panel}, {which contains a series of images and plots produced using data from CFHTLenS, as well as $J$ and $W1$ photometry, in order to confirm if the peaks correspond to real objects.}

{The left hand column of \autoref{fig:6panel} shows two images of the region in optical (top, from CFHTLenS) and in $W1$ (bottom, {\it WISE}). Cyan triangles mark the most likely cluster members, as these galaxies have colours consistent with the EzGal model, whilst green circles mark those that have colours inconsistent with the model, and are therefore less likely to be members. Orange symbols are galaxies within $10\arcsec$ of the provided centre coordinates for this object, and are candidates for the brightest central galaxies of the cluster. The top-centre panel shows a zoomed in version of the density map ($3.75\arcmin$ $\times$ $3.75\arcmin$), centred on the largest peak. The white plus signifies the coordinates of the object centre and the purple circle is the location of the peak. The top-right panel shows histograms of the redshifts of galaxies within the CFHTLenS catalogue (black), and a subset of those that also appear in the {\it WISE} data (green). The object's recorded redshift is marked with a red dashed line, and the photometric redshifts of the most central galaxies are marked with orange dashed lines. The bottom-centre panel shows a $J$-$W1$ colour-magnitude diagram, with the red dashed line indicating the colour predicted by the EzGal model for an L$^{\star}$ cluster galaxy at the object's recorded redshift. The red transparent band marks a fiducial $\pm0.2$ magnitude error on this value. The bottom-right panel shows colour against the CFHTLenS photometric redshift for galaxies in the colour-magnitude diagram, with identical red lines to the bottom-centre and top-right plots showing the expected colour for the object and the recorded photometric redshift.}

{The primary resource for identifying whether the object is indeed a true detection is using the top-right plot} and counting the number of peaks in the redshift distribution. In this case, using object SA6 which appears in \autoref{fig:6panel} as an example known lens, we would count two peaks, one at $z\simeq0.3$ and one at $z\simeq0.6$. We also stress that qualitatively the serendipitous detections near to SA6 are typical of those found in the outskirts of the maps. Note the broad distribution of $(J-W1)$ colour at the latter redshift, which arises due to our method in which we cope with the multiple possible matches for galaxies at the centre of dense cluster cores, occurring due to the PSF of {\it WISE} being distinctly larger than that of VISTA. In the absence of relying on galaxy colours, galaxies are matched based on the closest entry in the other catalogue, up to a maximum separation of 1.4 arcseconds. Multiple matches are therefore common in high density cluster cores where multiple J-band detections are separated by less than the {\it WISE} PSF. This leads to inconsistent colour measurements either because a $W1$-band flux measurement contains contributions from multiple individual $J$-band sources (leading to an overestimate of $J-W1$), or because galaxies in different wavebands are simply matched incorrectly. 

In total we examine $269$ serendipitous detections with SNR$_{\rm eff}\ge3$, of which $70$ have SNR$_{\rm eff}\ge5$. At SNR$_{\rm eff}\ge5$, $66$ detections are associated with one or more redshift peaks and therefore are classified as secure detections in that there is clear evidence of one or more collapsed group/cluster-scale halos along the line of sight. This translates to a false positive rate of $\simeq6$ per cent at SNR$_{\rm eff}\ge5$. At $3\le$ SNR$_{\rm eff}<5$ we find that $158/199$ serendipitous detections are associated with one or more redshift peaks, giving 45 total detections for SNR$_{\rm eff}\ge3$ that are not associated with redshift peaks. This implies a false detection rate of $\simeq17$ per cent at SNR$_{\rm eff}\ge3$. We also find no evidence of a strong trend in false positive rate at $3\le$ SNR$_{\rm eff}<5$, and expect an increase in false positive rate below SNR$_{\rm eff}=3$ due to a large increase in the number of detections. {To summarise, including SNR$_{\rm eff}>3$ detections only increases the false positive rate of detections to $17$ per cent whilst producing almost four times as many real detections than if we include only the SNR$_{\rm eff}>5$ detections.} In addition, when comparing to our tests on objects from our test sample (\autoref{sec:results_real}), extending the detection threshold down to SNR$_{\rm eff}\ge3$ increases the recovery of lenses with small Einstein radii, $3<\theta_{\rm E}\le5\,\rm arcsec$ from $\simeq40$ per cent to $\simeq70$ per cent (\autoref{fig:recov_frac_histogram}). This is encouraging for building a watchlist based on $J$ and $W1$-band photometry in advance of LSST, especially bearing in mind that photometric redshifts from LSST photometry and spectroscopic redshifts from surveys taking advantage of instruments such as 4MOST will be efficient tools to subsequently suppress the false positives. {Therefore, we believe SNR$_{\rm eff}>3$ to be a sufficient threshold for selecting detections of real objects to include in a lensed transient watchlist.}

\subsection{Testing on Rubin DP0 data}
\label{sec:dp0_test}

The Rubin Observatory Data Preview 0 (DP0) is a programme providing access to simulated LSST-like data and images. These data were produced following the Dark Energy Science Collaboration's (DESC) second data challenge \citep{Korytov19,DescDc2}, whereby the results of N-body simulations are passed through a model of the Rubin hardware before being processed by the official LSST pipelines, producing a catalogue of objects that is representative of what is to be expected from the Rubin telescope. It includes data for 300 square degrees of sky in all six Rubin bands and is based on stacked images containing five years of observation data. The objects within the simulation reach out to $z=3$, however not every object will necessarily be detected following the processing steps. The base simulated data set contains information on individual dark matter halos and the galaxies bound to them, presenting an opportunity to test our method on data where the true underlying distribution of dark matter halos is known.

We applied our method to a single 0.5$\times$0.5 region of the DP0 catalogue that was known to contain a significant number of massive and less-massive halos. We select galaxies by choosing extended objects that are detected in the $y$-band brighter than 21.1 mag, which corresponds to the $J$-band detection limit of VISTA corrected to the $y$-band using the expected $y$-$J$ colour for an L$^\star$ galaxy as calculated using the same EzGal model as described in section \autoref{sec:WISE}. Our selection is therefore tuned to be as similar as possible to the current detection limits of VISTA $J$-band, and makes for a reasonable comparison despite the DP0 dataset being significantly deeper. All other parameters used in creating and convolving maps are identical to those used with the VISTA and {\it WISE} data as described in \autoref{sec:method}. 

\begin{figure}
    \centering
    \includegraphics[width=1.0\columnwidth]{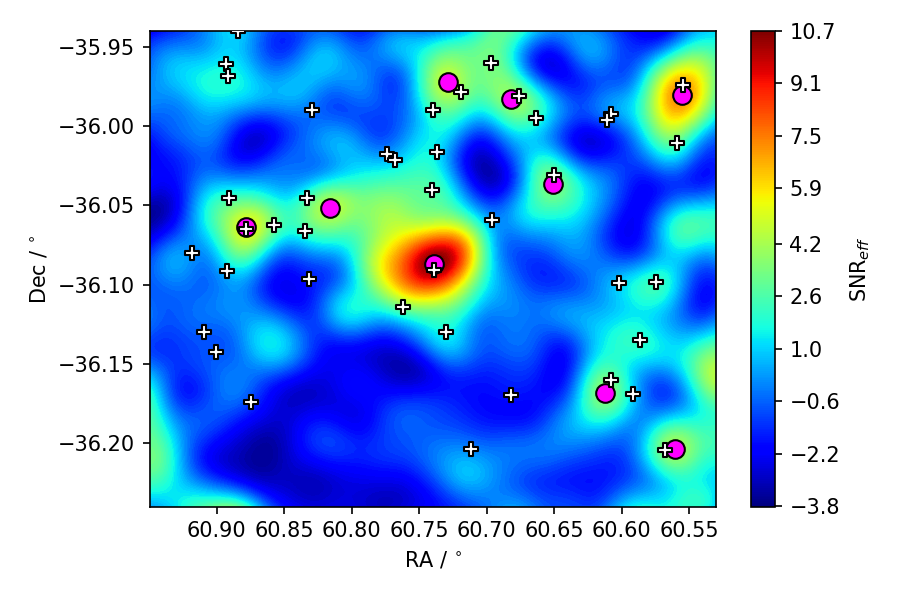}
    \caption{The map produced for the test region of simulated Rubin data. Magenta dots mark the locations of peaks with SNR$_{\textrm{eff}}>3$, and white pluses mark centres of dark matter halos with mass $M \geq 10^{13}$M$_{\odot}$. Qualitatively, the distribution of dark matter halos follows the high SNR$_{\textrm{eff}}$ regions of the map, which is indicative that the method is functioning as expected. The 9 peaks are all located within 1 arcmin of a dark matter halo, indicating the method has a low false positive rate when cutting at SNR$_{\textrm{eff}}=3$.}
    \label{fig:dp0map}
\end{figure}

\begin{figure}
    \centering
    \includegraphics[width=1.0\columnwidth]{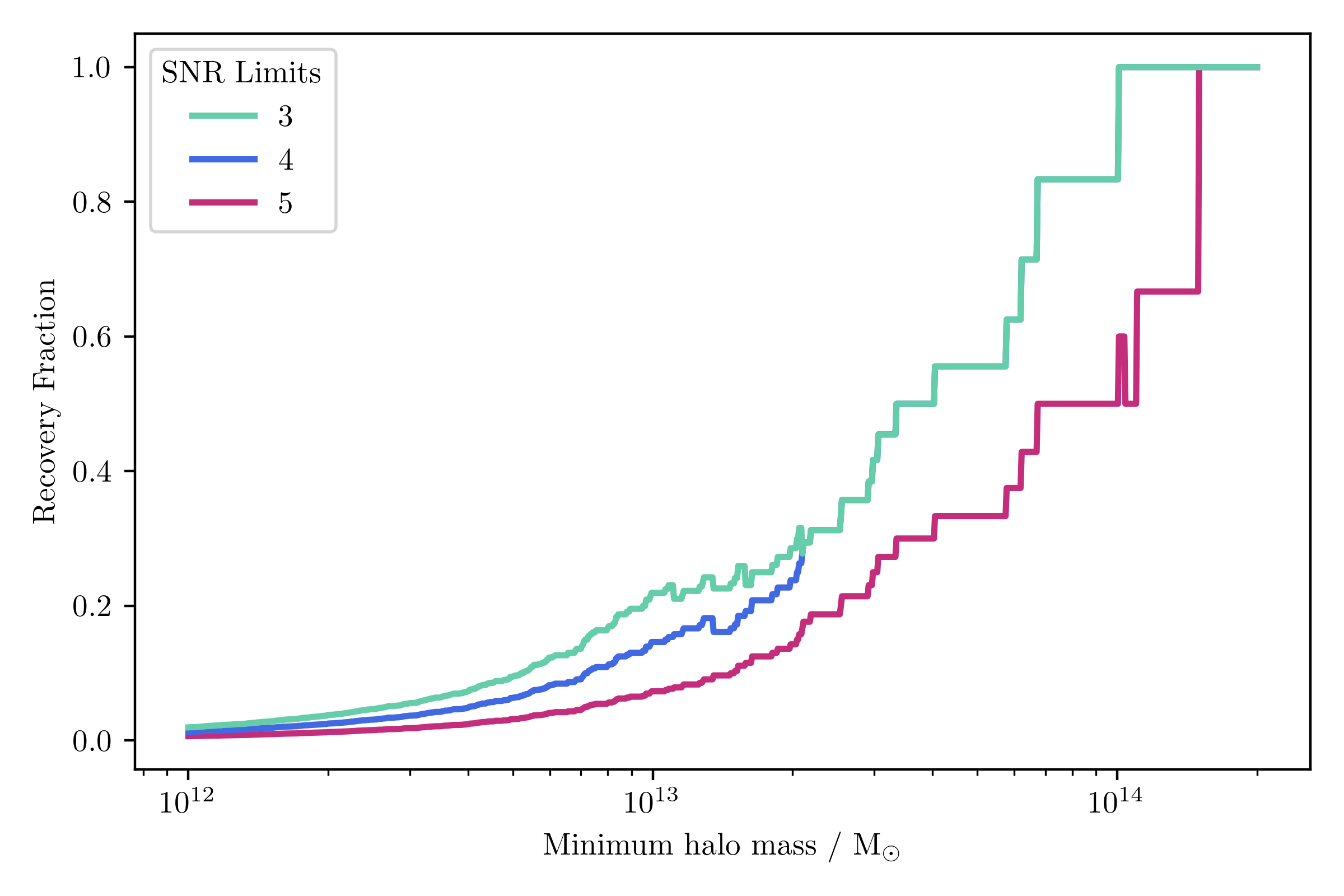}
    \caption{The recovery fraction of dark matter halos within the test region of simulated Rubin data as a function of halo mass. We calculate the fraction of halos greater than a given mass that are recovered by the method, where an object is considered to be recovered if it is the closest object within 1 arcmin of an SNR$_{\textrm{eff}}>3$, $4$ or $5$ peak (note that the SNR$_{\textrm{eff}} = 3$ and $4$ curves are identical above $M \sim 2\times10^{13}$M$_\odot$). The recovery rate using the simulated data matches well to the real data, with a good rate at high masses, that falls off at smaller values due to the difficulty in detecting those with fewer members. The ability to recover objects associated with massive dark matter halos does not explicitly match the goals of using the method, but it is sufficiently related, easy to test and gives a reasonable metric for the method's capability.}
    \label{fig:dp0recFrac}
\end{figure}

Within the simulation test region, there are 41 dark matter halos with $M \geq 10^{13}$M$_{\odot}$ at $z < 1$. We restrict to this redshift as the selection is tuned to the detection limits of $L^\star$ cluster galaxies in the $J$-band as shown in \autoref{fig:depths}, which we estimate to be detectable up to $z\simeq 1$. \autoref{fig:dp0map} shows the map for the test region, with the locations of the 41 dark matter halos overlaid. Firstly, it is reassuring to note that the qualitative distribution of these dark matter halos mostly follows the higher SNR$_{\textrm{eff}}$ regions of the map, which matches expectations. The map contains 9 peaks with SNR$_{\textrm{eff}}>3$, all of which are within 1 arcmin of at least one of these halos. Therefore none of these peaks can be considered a false positive detection, based on the definition established in the previous section, which further affirms and is consistent with our previous evaluation that the false positive rate of the method is low when cutting at SNR$_{\textrm{eff}}\geq3$. 

In order to quantify the recovery fraction of objects within the test region, we assume that the halo closest to each peak is the one which provides the greatest contribution to that peak's signal, and hence is the object considered to be recovered by the presence of the peak. In reality, any given peak arises from the contribution of many objects that may be related to multiple different halos, however this simplification allows us to relate the recovery to a single parameter -- the halo mass. We vary the minimum mass of the halos we consider, and plot the recovery fraction (as defined above) as a function of the halo mass in figure \autoref{fig:dp0recFrac}, {and for this we extend the lower mass cut for halos down to $10^{12}$M$_\odot$}. This shows that the recovery fraction is  $>80$ per cent for $M_{200}>7\times10^{13}\rm M_\odot$ clusters
and drops as we extend to include the less-rich lower mass halos which do not produce as strong of a signal in the map. 

\section{Summary}
\label{sec:summary}

{Watchlist-based searches for gravitationally lensed transients provide a promising route toward the discovery of many candidates including supernovae, kilonovae, gravitational waves and gamma-ray bursts from various wide-field survey data streams. The watchlist-based approach allows transients detected near watchlist objects to be quickly highlighted as candidates, although the lensing nature of these events would need to be confirmed by utilising further follow-up observations and/or analysis of the data, depending on the transient in question.}

{A wide-field survey of particular interest is the upcoming Rubin Observatory LSST, which is expected to find $\sim$millions of supernovae and $\sim$thousands of kilonovae during its operation -- some of which will be lensed. However, in order to maximise the opportunities for discovery of the rare lensed cases, and to enable science early in Rubin's operations, the watchlist must be assembled in advance. Sufficient data already exists in the $J$ (from VISTA/UKIRT) and $W1$ (from {\it WISE}) bands to produce such a watchlist that covers the entire sky, and is deep enough to include the majority of the lenses associated with the high-redshift population of lensed transients.} We have described a proof-of-concept for a method which utilises this data to discover galaxy groups and clusters all-sky out to $z\sim1$. In testing, our method successfully recovers between 80 and 100 per cent of a sample of large Einstein radius ($\theta_E \geq 5\arcsec$) galaxy clusters, and 40 to 70 per cent of smaller Einstein radius objects (dependant on individual radii and the SNR$_{\textrm{eff}}$ detection threshold used). We also estimated the false positive rate of the method by investigating the surrounding regions of test objects that the method identifies as likely to contain a group or cluster. By inspecting telescope images and by producing and inspecting both colour-magnitude diagrams and plots of the photometric redshift distribution of galaxies local to these candidates, we searched for evidence of real group or cluster objects at these locations. Using these, we were able to constrain the false positive rate for significant detections to be between approximately 6 to 17 per cent, depending upon the applied SNR$_{\textrm{eff}}$ detection threshold. 

Our investigations clearly show that the applied value of the SNR$_{\textrm{eff}}$ threshold changes both the level of completeness and the false positive rate of the method. Using a higher SNR$_{\textrm{eff}} = 5$ threshold results in secure detections of most large Einstein radius clusters with a very low false positive rate, but conversely produces a less complete catalogue due to many groups and clusters with smaller Einstein radii being detected with lower SNR$_{\textrm{eff}}$. We concluded from our testing that a lower SNR$_{\textrm{eff}} = 3$ threshold is optimal, as the increased number of recovered objects outweighs the increase in false positives. We also anticipate being able to suppress the number of false positives in the future once data from Rubin becomes available. 

We also tested the method on simulated data from the Rubin DP0 programme, that was selected to be comparable to currently available data sets. Testing on this data has the advantage of knowing the true underlying distribution of the dark matter halos that drive the formation of lenses in our universe. This test provides similar results, indicating a good recovery rate (>80\% for halos above $10^{14}$M$_\odot$) and {with no false positives produced in our test region.}

{Future work will focus on producing a catalogue of cluster detections across the full sky using the method described here, and will re-evaluate both recovery and false positive rates by comparison of other large cluster catalogues such as those assembled by \citet{redmapper14}, \citet{WenHan18} and \cite{Finoguenov20}, ensuring to collectively encompass multiple detection methods including optical/red sequence, galaxy overdensity and x-ray detection.}

\section*{Acknowledgements}
This publication makes use of data from observations obtained as part of the VISTA Hemisphere Survey, ESO Progam, 179.A-2010 (PI: McMahon), as well as data products from the Wide-field Infrared Survey Explorer, which is a joint project of the University of California, Los Angeles, and the Jet Propulsion Laboratory/California Institute of Technology, funded by the National Aeronautics and Space Administration.

DR (ORCID 0000-0002-4429-3429) acknowledges a PhD studentship from the Science and Technology Facilities Council. GPS acknowledges support from The Royal Society, and the Leverhulme Trust. GPS, MB and SM acknowledge support from the Science and Technology Facilities Council (grant number ST/N021702/1). MJ is supported by the United Kingdom Research and Innovation (UKRI) Future Leaders Fellowship (FLF), 'Using Cosmic Beasts to uncover the Nature of Dark Matter' (grant number MR/S017216/1).

\section*{Data Availability}
The VISTA and WISE data underlying this article are available from the VISTA science archive (\url{http://horus.roe.ac.uk/vsa/index.html}), and the NASA/IPAC infrared science archive (\url{https://irsa.ipac.caltech.edu/Missions/wise.html}), respectively. The sources of data corresponding to the samples of galaxy clusters are available through the relevant publications cited: \citet{Smith18}, \citet{CarrascoClusterSample}. Rubin DP0 data are not publicly available and are only accessible to those with Rubin data rights and that are DP0 delegates. More information can be found on the Rubin Observatory community forum (\url{https://community.lsst.org}).


\bibliographystyle{mnras}
\bibliography{bib}

\bsp	
\label{lastpage}
\end{document}